# Ideal two-dimensional electron systems with a giant Rashba-type spin splitting in real materials: surfaces of bismuth tellurohalides

Sergey V. Eremeev[1,2,3,*], Ilya A. Nechaev[2,3], Yury M. Koroteev[1,2,3], Pedro M. Echenique[3,4,5], and Evgueni V. Chulkov[3,4,5]

1. Institute of Strength Physics and Materials Science, pr. Academicheskiy 2/4, 634021, Tomsk, Russia
2. Tomsk State University, pr. Lenina 36, 634050, Tomsk, Russia
3. Donostia International Physics Center (DIPC), Paseo de Manuel Lardizabal, 4, 20018 San Sebastián/Donostia, Basque Country, Spain
4. Departamento de Física de Materiales, Facultad de Ciencias Químicas, UPV/EHU, Apdo. 1072, 20080 San Sebastián, Basque Country, Spain
5. Centro de Física de Materiales, CFM-MPC, Centro Mixto CSIC-UPV/EHU, Apdo.1072, 20080 San Sebastián/Donostia, Basque Country, Spain

**Spintronics is aimed at active controlling and manipulating the spin degrees of freedom in semiconductor devices. A promising way to achieve this goal is to make use of the tunable Rashba effect that relies on the spin-orbit interaction (SOI) in a two-dimensional (2D) electron system immersed in an inversion-asymmetric environment. The SOI induced spin-splitting of the 2D-electron state provides a basis for many theoretically proposed spintronic devices. However, the lack of semiconductors with large Rashba effect hinders realization of these devices in actual practice. Here we report on a giant Rashba-type spin splitting in 2D electron systems which reside at tellurium-terminated surfaces of bismuth tellurohalides. Among these semiconductors, BiTeCl stands out for its isotropic metallic surface-state band with the $\bar{\Gamma}$-point energy lying deep inside the bulk band gap. The giant spin-splitting of this band ensures a substantial spin asymmetry of the inelastic mean free path of quasiparticles with different spin orientations.**

* e-mail: eremeev@ispms.tsc.ru

The spin-orbit interaction (SOI) that causes spin-splitting of electron states in inversion-asymmetric systems [1–3] is expected to be efficiently exploited in spintronics. Even in the case of inversion-symmetric bulk materials, the spin-splitting induced by the SOI appears in two-dimensional (2D) geometries, such as, e.g., metal surfaces [4-6], metallic surface alloys [7,8], and semiconductor heterostructures [9,10]. This Rashba effect leads to shifting of opposite spin polarized bands by the momentum $k_R$ in opposite directions. In semiconductor 2D electron systems, due to the Rashba effect there is a possibility for controllable spin manipulation via an applied electric field [11–13], that provides a pathway to technology and miniaturization of spintronics devices. The key operating characteristic here is the magnitude of the SOI induced spin-splitting characterized by the Rashba energy of split states $E_R$ and the Rashba coupling parameter $\alpha_R = 2E_R/k_R$, which measures the strength of the spin-splitting.

For the conventional narrow-gap semiconductor structures, the parameter $\alpha_R$ is of order of $10^{-1}$ eVÅ [13]. Such a small Rashba splitting hampers the development of spintronics devices for room temperature applications since the latter require a significantly larger spin-splitting. Lately found systems that demonstrate a large spin-splitting of 2D-electron states ($α_R$ is about one order of magnitude greater than in the semiconductor 2D systems) are noble metal based Bi-surface alloys [8,14–16]. However, in this case metallic substrates prevent both the spin-splitting tuning by external electric field and making a surface spin signal detectable because of the large bulk current.

Apart from spintronics applications, the large Rashba splitting is strongly desired also for a semiconductor film that being sandwiched between an *s*-wave superconductor and a magnetic insulator can be used in a setup for creating and manipulating Majorana fermions for topological quantum computation [17]. In order to advance in the search of a semiconductor 2D system with large and tunable spin-splitting, we consider such a class of layered semiconductors as bismuth tellurohalides. Interest in these materials has been triggered by recently published results [18-20], which have revealed a giant Rashba-type spin-splitting of bulk states in BiTeI.

In the present paper, we examine two 2D geometries of bismuth tellurohalide compounds: Te- and haloid-terminated surfaces. On the basis of *ab-initio* calculations, we show that the 2D electron systems are formed at the Te-terminated surface of bismuth tellurohalides in electron surface states which split off from the bulk conduction band and, hence, inherit the giant spin-splitting and spin structure from the bulk states. These spin-split surface states provide unique quasiparticle properties of the respective 2D systems, which we analyze within the *GW* approximation. We show that among the considered bismuth tellurohalides the BiTeCl compound should be an ideal candidate for a very promising material for spintronics applications. This is demonstrated on the basis of the following factors: (i) the giant spin-splitting of the surface state, which is comparable with that in the surface alloys; (ii) the spatial localization of the state in the topmost three-layer block and its formation mainly by orbitals of the surface Te and the subsurface Bi atoms; (iii) the sufficiently large energy interval in the band gap, which contains both branches of the spin-split surface state with isotropic free-electron-like dispersion; (iv) the spin asymmetry of the inelastic mean free path of quasiparticles in different branches, which can be tuned by changing both the spin-orbit interaction strength via applying external electric field and the chemical potential.

**RESULTS**
**Bulk electronic structure of bismuth tellurohalides.** The bismuth tellurohalides BiTeCl, BiTeBr, and BiTeI have a hexagonal crystal structures [21] (see Fig. 1). BiTeI with the space group *P*3*m*1 is composed by Bi, Te, and I layers stacking along the *c* axis. BiTeBr has a $P\bar{3}m1$ space group; its structure differs from that of BiTeI by tellurium and bromine atoms randomly distributed within the Te and Br atomic layers. In BiTeCl ($P6_3mc$), the tellurium and chlorine layers alternate along the *c* direction of the unit cell, forming a four-layered packing [21]. All these compounds are characterized by ionic bonding and pronounced three layered (TL) structure, where the distance between TL's is about one and a half times greater than interlayer distances within the Te-Bi-*X* (*X*=Cl,Br,I) TL.

In Fig. 1, we show the bulk band structures calculated by VASP and FLEUR codes (see Methods) for the considered bismuth tellurohalides. As is seen from the figure, VASP gives almost the same band structures as FLEUR. In the case of BiTeI, the obtained electronic bands are in good agreement with earlier WIEN-code calculations [19]: both the conduction band minimum (CBM) and the valence band maximum (VBM) demonstrate giant Rashba-type spin splitting with $k_R \approx 0.055$ Å$^{-1}$ in the vicinity of the A point (the values of the Rashba coupling parameter $\alpha_R$ for CBM in the AH and AL directions are shown in Fig. 1). However, our calculated $\alpha_R$ is larger than that obtained in Ref. [19] that can mainly be attributed to larger $E_R$. It is worth noting that the FLEUR calculations performed without recourse to linearization error corrections (see Methods) give $\alpha_R \approx 4.8$ eVÅ that is close to $\alpha_R \approx 4.5$ eVÅ found in [19].

In BiTeBr, considered in ordered BiTeI-type structure, the Rashba-type splitting of bulk CBM and VBM bands decreases along with the increase of the band gap. In BiTeCl, the folding of the Brillouin zone along $k_z$ direction leads to transfer of the band gap to the Γ point. As one can see, the lighter *X* atom leads to widening of the band gap and to reduction of $\alpha_R$. In all the considered compounds the top of the valence band is mostly composed of X atom states while the bottom of the conduction band is formed by Bi orbitals. Te orbitals contribute to both gap edges. Note that ΓK/ΓM anisotropy of spin-splitting in BiTeCl is the lowest one among the considered compounds. The spin-resolved constant energy contours (CEC) for BiTeCl (Fig. 1(h)) demonstrate almost perfect circular shape for both inner and outer branches of the Rashba-split conduction band. This feature is distinct from the calculated CEC for BiTeI Ref. [18], which shows a hexagonal shape for the outer branch of the Rashba-split conduction band. This deviation of CEC from the circular shape is accompanied by sizeable out of plane spin component $S_z$ in BiTeI [18] while in BiTeCl the Rashba-split state has $S_z$ close to zero (not shown). Along with the marked differences the Rashba-split conduction band in BiTeCl has the same spin helicity as in BiTeI [18], i.e. clockwise spin rotation in the outer branch and counter-clockwise one in the inner branch.

The surface of BiTeCl and BiTeI formed under cleavage between neighboring TL can have two possible terminations: Te-layer termination or Cl(I)-layer termination, while in BiTeBr the only Te/Br mixed layer termination can be realized. We do not consider the latter surface in the present study.

**Surface band structure of BiTeCl.** To simulate semi-infinite BiTeCl(0001) with Te-terminated surface we passivated the Cl termination of the 8 TL slab by a hydrogen monolayer. The electronic band spectrum of this slab is shown in Fig. 2(a), where red and gray circles denote weights of the states localized in opposite, Te and H-Cl-terminations of the slab, respectively. The magnified view of the Te-terminated BiTeCl(0001) surface band structure near the minimum of the conduction band is given in Fig. 2(b). As is clearly seen from the figure, at the Te-terminated surface the spin-orbit split surface state arises well below the conduction band and is spatially localized in the topmost TL. The state is mainly formed by orbitals of the surface Te and subsurface Bi atoms (Fig. 2(c,d)). Within the energy gap region, the energy dispersion of the surface state demonstrates the free-electron-like parabolic character. The spin-splitting of the state in the $\overline{\Gamma} - \overline{M}$ is characterized by the spin-orbit interaction strength $\alpha_R$ = 1.78 eVÅ.

The spin-resolved CEC for the Rashba-split surface states on Te-terminated BiTeCl(0001) at -200 and -300 meV demonstrate circular shape in the gap region with the spin polarization inherited from the bulk conduction band. Similar to the states in the bulk conduction band, the surface state is almost completely in-plane spin-polarized but the $S_z$ spin component increases slightly when approaching the conduction band bottom.

The emergence of the surface state can be understood from analysis of changes of the near surface potential. In Fig. 2(e) the change of the local potential averaged over $xy$ planes with respect to that in the central part of the slab is shown. Owing to the pronounced TL structure of the compound, the observed $\Delta V$ is stepwise. The largest decrease of the potential is in the outermost TL, while in the third TL it is negligible. Trapping of electrons of the outermost TL in the surface potential well splits off this Rashba surface state from the bulk conduction band and noticeably increases its effective mass. Additionally, the presented $\Delta V$ reflects modifications of the potential within the outermost TL, which lead to decreasing the strength $\alpha_R$ with the unchanged atomic contribution to the latter.

To complete the picture of surface properties of BiTeCl, we examine its Cl-terminated surface. As in the previous case it is simulated by the 8 TL slab with hydrogen on opposite (Te-terminated) side of the slab. In contrast to the Te-terminated surface, on the Cl-terminated BiTeCl(0001) the positive $\Delta V$ (which is two times larger by magnitude than that on the Te-terminated surface (see Fig. 3(a)) splits off the valence band edge states, forming in the gap a pair of spin-split surface states with negative effective masses (Fig. 3(b)). The resulting surface structure resembles dispersion of spin-orbit split surface states in surface alloys (see, e.g., [16]). Owing to noticeable anisotropy of the surface state dispersion of the upper state the CEC shown in Fig. 3(c) has more complex shape. Another feature of the surface states on Cl-terminated surface is that they are appreciably out-plain spin polarized. Due to above mentioned features and the fact that under the degeneracy point the branches of the spin-split states are practically in immediate proximity to each other this Cl-terminated surface has less appeal than the Te-terminated one.

**Quasi-particle dynamics.** Apparently, the revealed electronic structure of the Te-terminated surface of BiTeCl represents a unique case that combines the giant spin-splitting characteristic of noble metal based Bi-surface alloys and the controllability of semiconductor low-dimensional systems. Various chemical and physical phenomena at such a surface may be determined by decay of elementary excitations on the surface. In order to gain insight into quasiparticle dynamics at the Te-terminated surface of BiTeCl, we consider here the inelastic decay caused by electron-electron scattering in the 2D electron system formed by electrons in the surface state.

Due to the fact that the surface-state wave function is mostly confined within the topmost TL [see Fig. 2(d)], and that the $\bar{\Gamma}$-point energy of this surface state lies deep inside the bulk band gap, we suppose that to a certain extent the respective 2D electron system can be an isolated one. As a consequence, the behavior of quasiparticles in such a system can be adequately described within the $G^0W^0$ scheme of Ref. [22] based on the Rashba model. In this scheme, the considered 2D electron system is modeled by a 2DEG with the Rashba SOI, where the Coulomb interaction is reduced by a factor which is equal to the dielectric constant $\varepsilon_\infty$ of the medium, where the electron system resides, i.e., of the bulk BiTeCl (see Methods). By performing $G^0W^0$ calculations, we examine hole and electron excitations in the energy region that does not contain bulk-projected states.

The $G^0W^0$ results obtained with different values of the Fermi energy ($E_F$) in the mentioned 2D system are shown in Fig. 4. Since the chemical potential can be continuously tuned by doping or applied electric field, the considered locations of the Fermi level are realizable in practice. The results presented in Fig. 4 predict unique peculiarities for the considered 2D system. The first one is a noticeable difference between the inelastic decay rates $\Gamma_{inner}$ and $\Gamma_{outer}$, which is observed in the electron excitations energy region. This difference means that the inelastic decay rate depends on the branch of the spin-orbit split band, and, as a consequence, on spin orientation for a given direction of **k**. The ratio $\Gamma_{outer}/\Gamma_{inner}$ demonstrates dependence on energy, which varies strongly with $E_F$. The ratio reflects a spin asymmetry of the decay rate and, owing to equal quasiparticle velocities at a given energy for both branches, characterizes the ratio of the corresponding inelastic mean free paths (IMFP) of electrons carrying different spin. The resulting IMFP spin asymmetry is essentially bigger [especially in the cases shown in Figs. 4(c) and 4(d)] than in 2D electron systems with small $α_R$ and amounts to values which are already well comparable with those in ferromagnets (see, e.g., Ref. [23]). However, as distinct from ferromagnetic materials, in our case the values of the IMFP spin asymmetry can be tuned by external electric field that modifies $α_R$.

As regards the mentioned spin asymmetry, the most interesting case is that shown in Fig. 4(d), where within the energy interval of ~60 meV the ratio, being measured from unity, changes its sign. This effect is caused by opening the plasmon decay channel for inter-branch transitions with quite small momenta and finite energies. The same reason leads to appearance a "loop" in the energy dependence of $\Gamma_{outer}$ in the hole excitations region below the degeneracy point $E_0$ [see Figs. 4(a) and 4(b)]. This is the second unique peculiarity of the considered 2D electron system, which can easily be resolved experimentally due to the revealed giant spin splitting of the surface state.

The revealed behavior of the inelastic decay rate can be observed by analyzing the linewidth of momentum distribution curves (MDC) measured by angle-resolved photoemission spectroscopy (ARPES). Actually, an MDC, being a $k_∥$-cut through the quasiparticle spectral function at a fixed energy, demonstrates peaks at the momentum values which correspond to the dispersion of the branches, whereas height and width of these peaks are related to the inelastic decay rate. Fig. 4 represents both the spectral function and the modeled MDC. As is seen from Figs. 4(a) and 4(b), in the case of appearance of the plasmon decay channel for holes the left peak is expected to be lower and wider than the right one. In the case of $E_F = E_0$ [Fig. 4(c)], where there is no plasmon channel for holes, the peaks are in the inverse ratio, and as a whole the situation with hole excitations becomes similar to that considered in Ref. [24] as a hypothetical case. Being experimentally observed, the mentioned feature can be evidence that the system that is formed by electrons in the surface state is a sufficiently isolated 2D system as we suggested above.

As to the conventional plasmon decay channel that appears in the electron excitation energy region, the contribution of such a channel to the decay rate no longer has the familiar sharp form as in a 2D electron system without the SOI. This form gets smooth due to the finite plasmon linewidth originated from the extension of the Landau damping region (the region where plasmons decay into single-particle excitations). The extension is caused by appearance of inter-branch transitions and in the case of $E_F = E_0$ becomes such wide that the plasmon spectrum entirely lies in the SOI-induced damping region.

**Surface band structure of BiTeI.** Now we turn to the Te-terminated surface of BiTeI. Such a surface also holds the Rashba-split surface state (see Fig. 5(a)) well reproducing the ARPES data [18]. The calculated Rashba coupling parameter $α_R$ = 3.5 eVÅ is in good agreement with the value of 3.8 eVÅ reported in Ref. [18]. In contrast to BiTeCl, the surface state of BiTeI is less split off from the bulk conduction band and penetrates deeper into subsurface TL's (Fig. 5(b)). In accordance with the experimental finding [18], the CEC of the surface state have circular shape for both inner and outer branches at the Fermi level, however below the conduction band bottom the outer branch CEC has hexagonal shape (Fig. 5(c)) and thus it can not be precisely described by the isotropic Rashba model. Such a hexagonal deformation leads to increasing $S_z$ spin component, i.e. to larger deviation from the in-plane spin polarization of the surface than it occurs in the case of chloride.

As regards quasiparticle properties in the case of BiTeI, the part of the surface state, which does not overlap with bulk-projected states [Fig. 5(a)], contains only the outer branch well below the degeneracy point. To avoid implication of the bulk-projected states in decay processes of hole and electron excitations, we put $E_F$ measured from $E_0$ at -80 meV. In this case, we have the desired energy range of ±36 meV with regard to the Fermi level. Within this range, the obtained energy dependence of $Γ_{outer}$ does not exhibit any specific behavior and in many respects repeats the $Γ_{outer}$ energy dependence shown in Fig. 4(d) below and slightly above $E_F$.

**DISCUSSION**

Thus, we have scrutinized the bulk and surface band structures of bismuth tellurohalides. We have shown that the giant spin-splitting of the surface states is inherited from that of bulk states. The surface states emerge by splitting off from the bulk conduction (at Te-terminated surfaces) or valence (at haloid-terminated surfaces) bands owing to modifications in the potential within the near-surface layers. By the example of BiTeCl, we have ascertained that in the case of a haloid-terminated surface the amplitude of these modifications is noticeably larger than that for a Te-terminated surface. As a result, the latter has one spin-split surface state in the energy gap only, whereas the haloid-terminated surface is characterized by a pair of spin-split surface states with negative effective masses. We have found that the 2D spin-orbit coupled electron system that can be formed by surface state electrons hosted by the Te-terminated surface of BiTeCl possesses large spin asymmetry of the inelastic mean free path of electron excitations. We believe that our findings will stimulate further theoretical and experimental investigations of layered polar semiconductors as materials which meet the requirements for successful spintronics applications. They provide desired electronic surface properties for efficient spin manipulation that is the basic ingredient of spintronics device functionality.

**METHODS**

For structural optimization and electronic bands calculations, we employed density functional theory (DFT) with the generalized gradient approximation (GGA) of [25] for the exchange-correlation (XC) potential as implemented in VASP [26,27]. The interaction between the ion cores and valence electrons was described by the projector augmented-wave method [28,29]. The Hamiltonian contained the scalar relativistic corrections, and the spin-orbit coupling was taken into account by the second variation method [30]. Complementary calculations of bulk electronic bands were performed using the full-potential linearized augmented plane-wave (FLAPW) method as implemented in the FLEUR code [http://www.flapw.de] within the GGA for the XC potential. The FLAPW basis has been extended by local orbitals of [31,32] to treat quite shallow semi-core *d*-states. Additionally, to reduce linearization error and accurately describe unoccupied states [33], we have also included for each atom one local orbital per angular momentum up to $l = 3$. For the VASP simulation of the "semi-infinite" BiTe*X* crystals, we used 8 TL slabs that were terminated on one side by a monolayer of hydrogen to clearly separate the surface states localized on opposite sides of the slab. In order to estimate the dielectric constant $\varepsilon_\infty$ for BiTeCl, first, with the use of the VASP code we have performed *ab-initio* calculations of $\varepsilon_\infty$ of BiTeI and BiTeCl within the random phase approximation (RPA) without taking the SOI into account. The exclusion of the SOI from consideration is caused by the fact that calculations of charge response for systems with the SOI are not a widespread option of *ab-initio* codes so far. Next, we have multiplied the obtained $\varepsilon_\infty = 11.53$ for BiTeCl by the ratio between the experimental ($\varepsilon_\infty = 19\pm2$ [34]) and calculated ($\varepsilon_\infty = 14.26$) values of the dielectric constant of iodide. As a result, we have ended up with the estimation $\varepsilon_\infty \approx 15.0$ for the chloride.

**Acknowledgements**

We acknowledge partial support by the University of the Basque Country (project GV-UPV/EHU, grant IT-366-07) and Ministerio de Ciencia e Inovación (grant FIS2010-19609-C02-00). E.V.C. thanks J. Fabian for enjoyable discussions.

**Author Contributions** First-principles calculations were performed by S.V.E., Yu.M.K., and I.A.N. $G^0W^0$ calculations were done by I.A.N. Figures were produced by S.V.E., and I.A.N. All authors contributed to the discussion and writing the manuscript. E.V.C. directed the project.

**Competing Interests** Reprints and permissions information is available at www.nature.com/reprints. The authors do not have competing financial interests.

**FIGURE LEGENDS**

**Figure 1 Bulk atomic and electronic structure of bismuth tellurohalides.** Atomic crystal structure (a-c), Brillouin zone of the hexagonal cell (d), and bulk band spectra of BiTe$X$ ($X$=I,Br,Cl) compounds (e-g) calculated by the VASP (red solid lines) and the FLEUR code (blue dashed lines) along high symmetry directions of the Brillouin zone. Band structure for BiTeBr calculated for BiTeI-type ordered phase. Spin-resolved constant energy contours for conduction band states in BiTeCl at 400 meV (h).

**Figure 2 Electronic structure of Te-terminated BiTeCl(0001).** (a) Band structure of a 8TL thick BiTeCl(0001) slab with hydrogen on Cl-terminated side; the red(gray) bands are states from the Te(H)-terminated side of the slab. The projected bulk band structure is shown in olive green. (b) A magnified view of electronic structure of Te-terminated BiTeCl(0001) surface in the vicinity of $\bar{\Gamma}$ [$E(k_\parallel)$ ranges correspond to blue dashed frame marked in the panel (a)]. Spatial distribution of the Rashba-split state charge density in the ($11\bar{2}0$) plane (c) and integrated over ($x$; $y$) planes (d). The change of the potential in near-surface layers of the crystal with respect to that in central, bulk-like, layers (e). Spin structure of the Rashba-split states on Te-terminated BiTeCl(0001), as given by spin projections $S_x$, $S_y$, and $S_z$ at energies of -200 meV (f) and -300 meV (g).

**Figure 3 Electronic structure of Cl-terminated BiTeCl(0001).** (a) The change of the potential in nearsurface layers of the crystal with respect to that in central layers for Cl-terminated BiTeCl(0001). (b) Band structure of Cl-terminated surface. (c) Spin structure of the Rashba-split states at energy of -80 meV.

**Figure 4 Surface-state quasiparticle dynamics at the Te-terminated surface of BiTeCl.** Left side of each panel: the inelastic decay rate for the inner and the outer branch ($\Gamma_{inner}$ and $\Gamma_{outer}$) and the ratio $\Gamma_{outer}/\Gamma_{inner}$ for electron excitations as functions of energy. Right side of each panel: contour plot of the respective quasiparticle spectral function and the momentum distributed curve (MDC) in arbitrary units and in the scale that is unified for all the panels presented. Energy is measured from the degeneracy point $E_0$. Values of the Fermi energy $E_F$ as well as the energy at which the MDC has been calculated are indicated.

**Figure 5 Electronic structure of Te-terminated BiTeI(0001).** (a) Surface band spectrum of Te-terminated BiTeI(0001). (b) Integrated over ($x$; $y$) planes charge density of the Rashba-split surface state. (c) Spin structure of the surface state at energy of the bulk conduction band bottom ($\approx$ -200 meV).

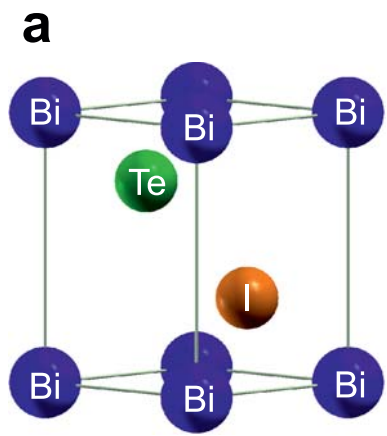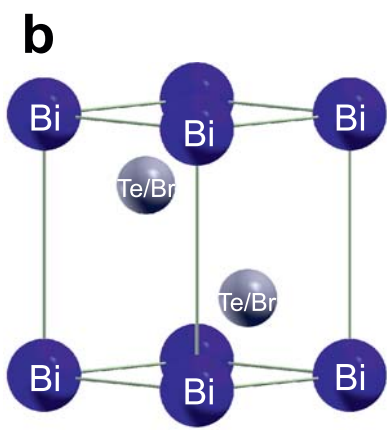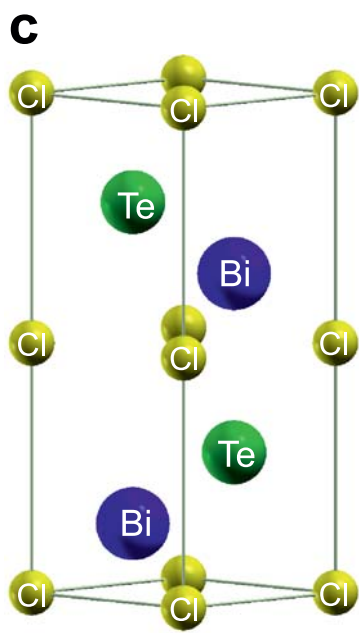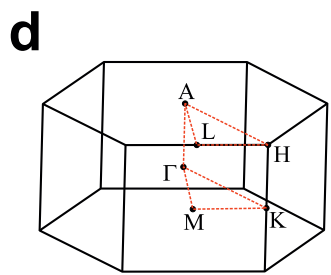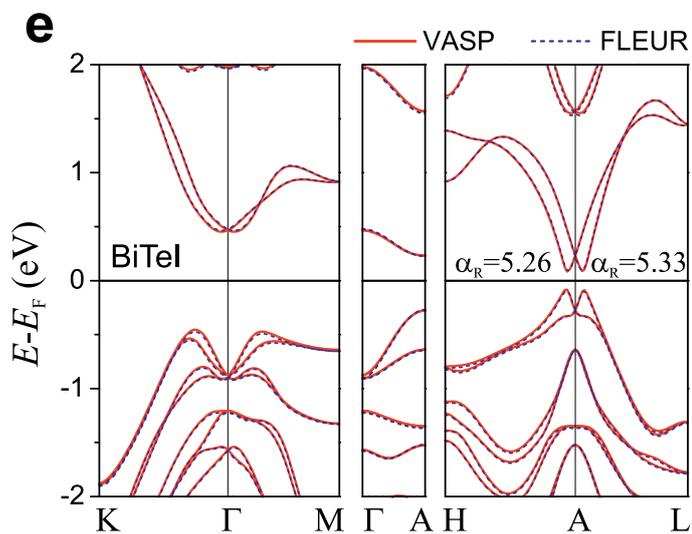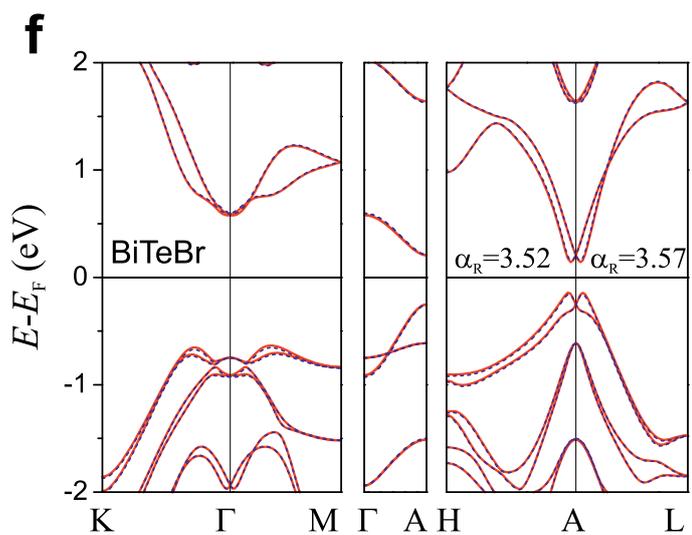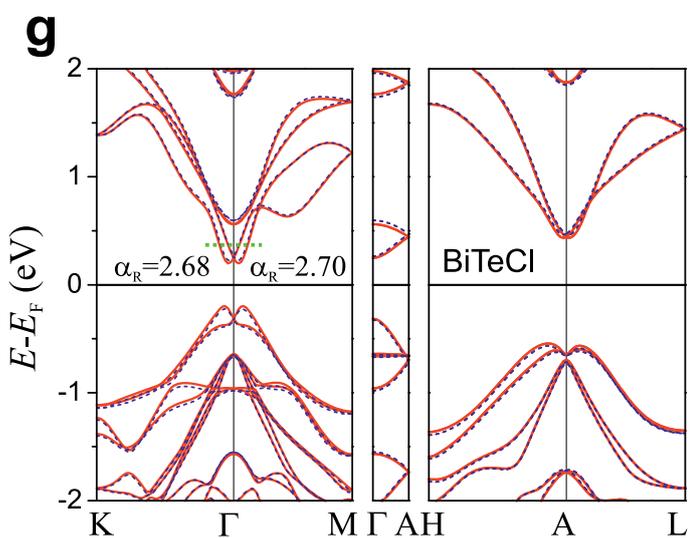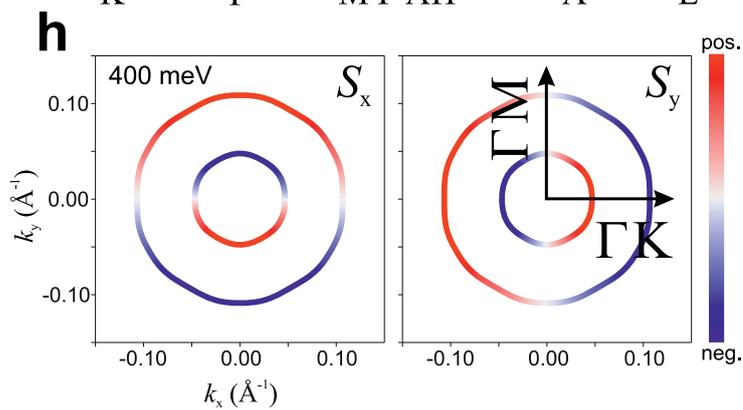

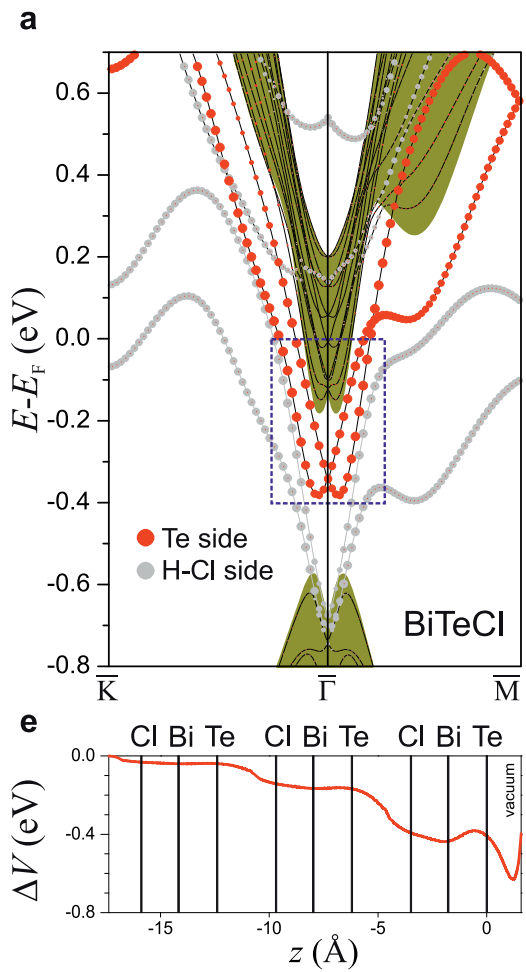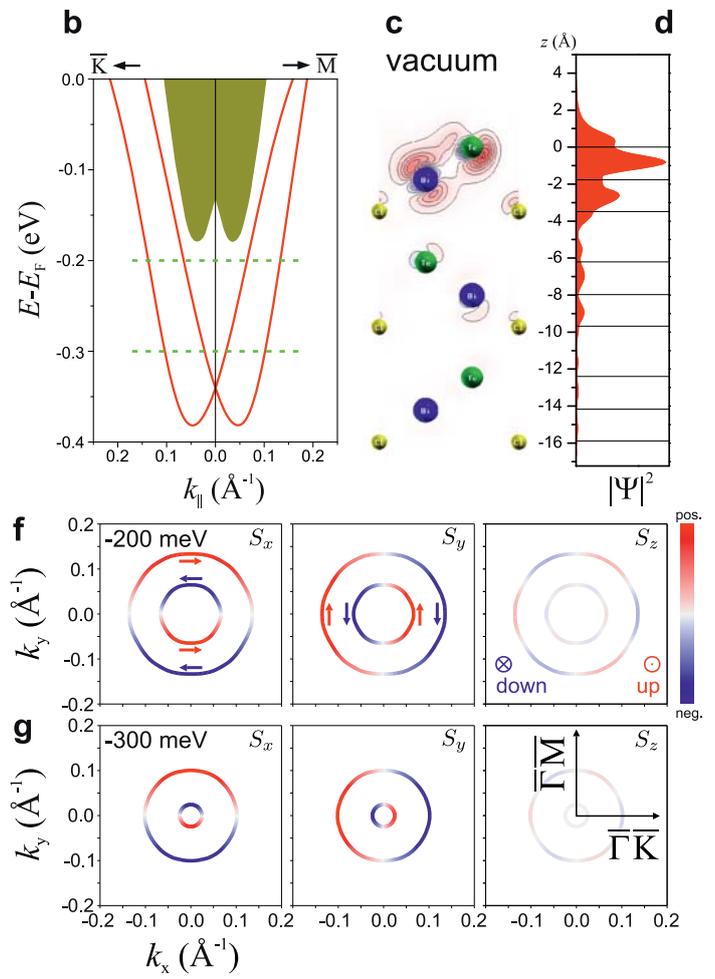

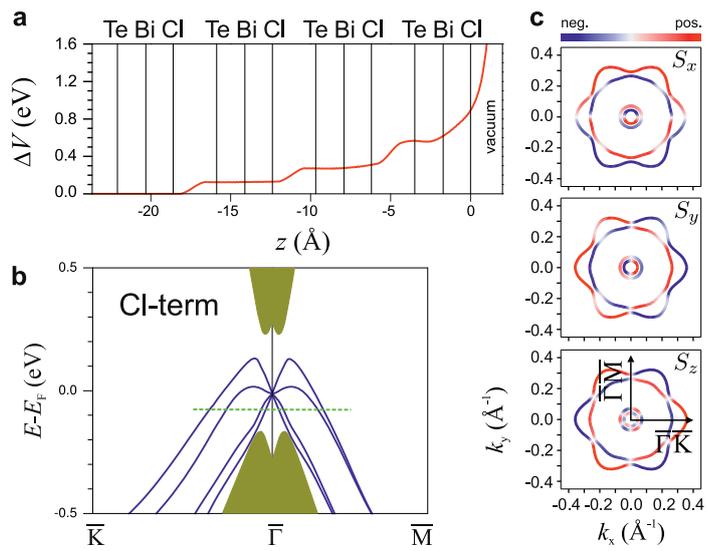

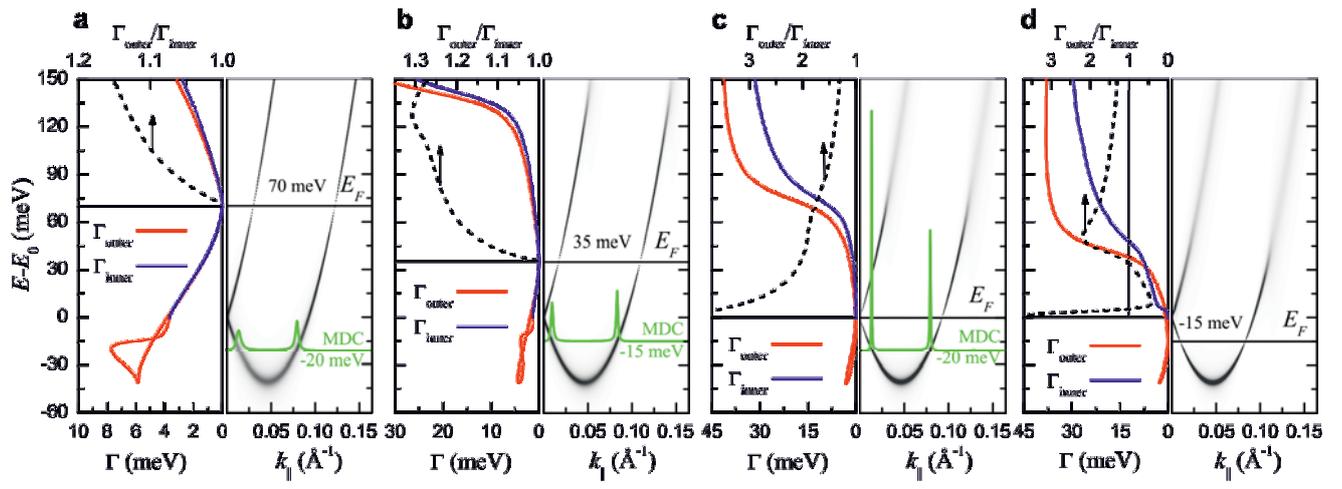

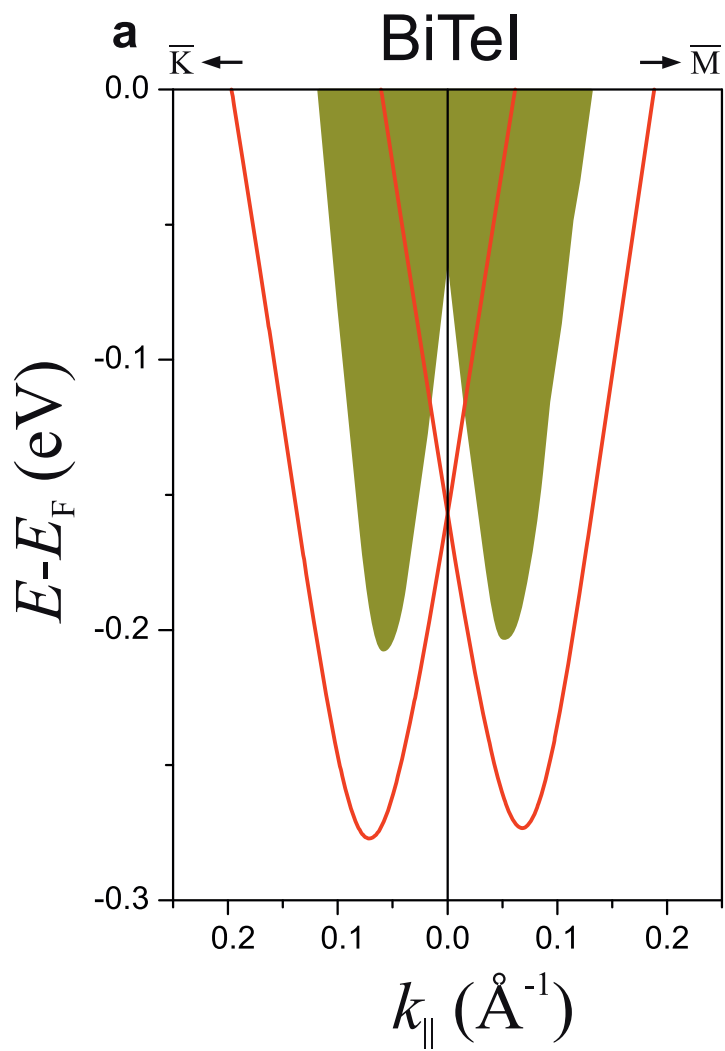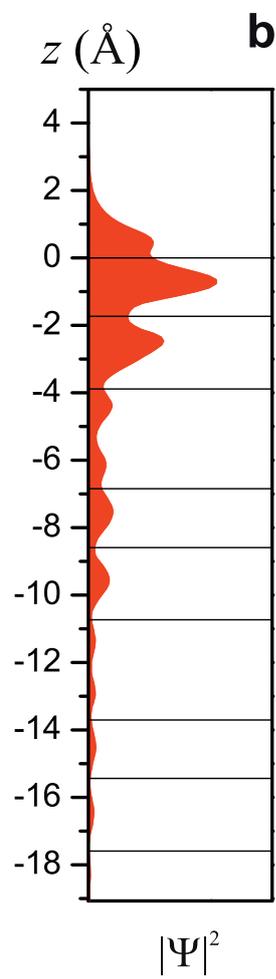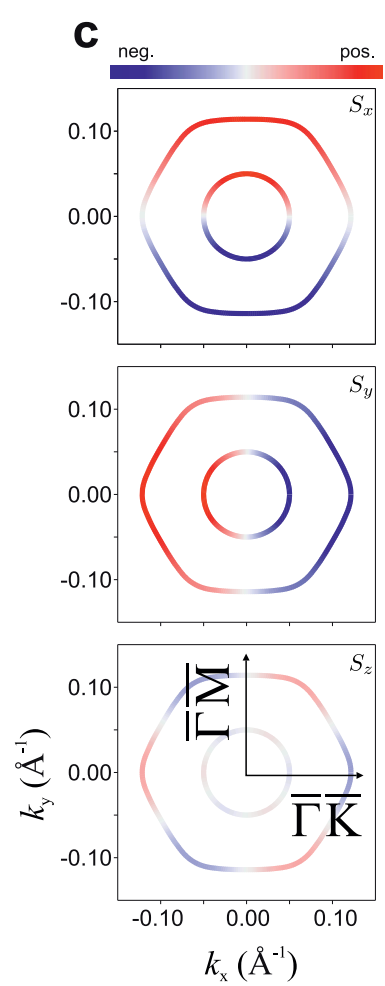